\documentclass[sigconf]{acmart}
\usepackage{currfile}
\usepackage{xstring}
\usepackage{listings}
\usepackage{xcolor} % Load the color package
\usepackage{algorithm}
\usepackage{algorithmic}
\usepackage{graphicx}
\usepackage{subfig}
\usepackage{lipsum}
\usepackage{enumitem,kantlipsum}
\usepackage{hyperref}
\usepackage{circuitikz}
\usepackage{soul}
\usepackage{array} % For centering text in tables
\usepackage{colortbl} % For coloring table cells
\usepackage{fancyhdr}

\AtBeginDocument{% 
  \providecommand\BibTeX{{%
    \normalfont B\kern-0.5em{\scshape i\kern-0.25em b}\kern-0.8em\TeX}}}

\copyrightyear{2025}
\acmYear{2025}
\setcopyright{rightsretained}
\acmConference[ASPDAC '25]{30th Asia and South Pacific Design Automation Conference}{January 20--23, 2025}{Tokyo, Japan}
\acmBooktitle{30th Asia and South Pacific Design Automation Conference (ASPDAC '25), January 20--23, 2025, Tokyo, Japan}\acmDOI{10.1145/3658617.3697764}
\acmISBN{979-8-4007-0635-6/25/01}

\newcommand{\bluehighlight}[1]{\colorbox[HTML]{dae8fc}{#1}} % Highlighting for the sections 
\newcommand{\lightredhighlight}[1]{\colorbox[HTML]{ffe6e6}{#1}} % Light red highlight
\newcommand{\deeperredhighlight}[1]{\colorbox[HTML]{ff8080}{#1}} % Deeper red highlight
\newcommand{\darkestredhighlight}[1]{\colorbox[HTML]{ff1a1a}{\textcolor{black}{#1}}} % Darkest red highlight
\newcommand{\greenhighlight}[1]{\colorbox[HTML]{1aff1a}{#1}} % Green highlight

\begin{document}

\title{K-Gate Lock: Multi-Key Logic Locking Using Input Encoding Against Oracle-Guided Attacks}

 \author{Kevin Lopez}
 %\orcid{0009-0002-3149-1968}
 \affiliation{%
 	\institution{ \textit{Computer Engineering \& Computer Science Department
 			\and California State University, Long Beach} }
 	\streetaddress{}
 	\city{}
 	\state{}
 	\country{}
 Kevin.LopezChavez01@student.csulb.edu
 }

 \author{ Amin Rezaei }
 % \orcid{0000-0002-7469-3642}
 \affiliation{%
 	\institution{ \textit{Computer Engineering \& Computer Science Department
 			\and California State University, Long Beach} }
 	\streetaddress{}
 	\city{}
 	\state{}
 	\country{}
 Amin.Rezaei@csulb.edu
 }

\begin{abstract}
Logic locking has emerged to prevent piracy and overproduction of integrated circuits ever since the split of the design house and manufacturing foundry was established. While there has been a lot of research using a single global key to lock the circuit, even the most sophisticated single-key locking methods have been shown to be vulnerable to powerful SAT-based oracle-guided attacks that can extract the correct key with the help of an activated chip bought off the market and the locked netlist leaked from the untrusted foundry. To address this challenge, we propose, implement, and evaluate a novel logic locking method called K-Gate Lock that encodes input patterns using multiple keys that are applied to one set of key inputs at different operational times. Our comprehensive experimental results confirm that using multiple keys will make the circuit secure against oracle-guided attacks and increase attacker efforts to an exponentially time-consuming brute force search. K-Gate Lock has reasonable power and performance overheads, making it a practical solution for real-world hardware intellectual property protection.
\end{abstract}




\maketitle

\section{Introduction}
\label{Sec:Intro}
In this split of design and manufacturing, one company designs the digital design, while another handles the physical fabrication of the Integrated Circuit (IC). While this separation of tasks poses a threat to chip security, logic locking \cite{EndingPiracy, 1fault_analysis} has emerged as a promising solution to prevent piracy and overproduction of hardware Intellectual Properties (IPs). Formally speaking, logic locking is the process of adding additional inputs to an IC, called key bits, to prevent the correct operation of the IC when the incorrect key is provided to the circuit. Traditionally, locking has been done using only one global key, which made it susceptible to the SAT-based attack \cite{evaluatingLogicEncryptionAlgorithms} that extracts the key using an oracle (i.e., a working chip bought off the market) and a locked netlist leaked from an untrustworthy foundry. While there have been attempts to reduce the success of the SAT-based attack to brute-force \cite{AntiSAT, SARLock}, sophisticated attacks have been proposed to find out the correct key of these methods.
    
We believe that the vulnerabilities associated with a single static key can be effectively mitigated through multi-key logic locking. In this paper, we introduce an advanced multi-key approach called \textbf{K-Gate Lock} where the inputs of each gate are encoded with different key values. To activate the circuit correctly, these values must be provided in the specific sequence used during the encoding process. In this paper, we present the following contributions: 
\begin{enumerate}
    \item [$\bullet$] Proposing a robust multi-key logic locking based on input encoding, implemented fully in combinational logic;
    \item [$\bullet$] Implementing an efficient algorithm to lock a circuit with multiple user-defined keys with tunable time complexity;
    \item [$\bullet$] Generating more than 40 benchmarks based on the proposed method, measuring the overhead, and evaluating its security against state-of-the-art oracle-guided attacks.
\end{enumerate}

\section{Background and Related Work}
\label{Sec:Related}
In this section, we first consider the evolution of logic locking techniques as well as oracle-guided attacks, and then review the existing efforts in multi-key logic locking.

\subsection{Logic Locking Techniques}
\label{Sec:Defenses}
Initial techniques of logic locking rely on single-key schemes, primarily employing {\sc xor}-based and {\sc mux}-based mechanisms \cite{EndingPiracy, 1fault_analysis}. In {\sc xor}-based logic locking, the key bits are matched with random inverters and buffers. Then, the {\sc xor} gates controlled by key bits are used to replace selected buffers and inverters. Additionally, {\sc mux}-based logic locking selects random wires and substitutes them with 2-1 {\sc mux}s whose inputs are real signals and random dummy ones, and selectors are the key bits. However, advancements in SAT solvers have been utilized to expose vulnerabilities in these methods \cite{evaluatingLogicEncryptionAlgorithms}, leading to the development of more robust techniques such as Anti-SAT \cite{AntiSAT}, SAR-Lock \cite{SARLock}, TT-Lock \cite{TTlock}, CAS-lock \cite{CAS-Lock}, BLE \cite{rezaei2020rescuing}, DLE \cite{6_Distributed_logic_encryption}, Full-Lock \cite{kamali2019full}, Cross-Lock \cite{shamsi2018cross}, HLock \cite{HLock}, TraceLL \cite{TraceLL}, TriLock \cite{TriLock}, and others \cite{7_Global_attack, 8_Hybrid_memristor, 9_Sequential_logic_encryption, 10_CoLA, CycSAT_cyclicLogicLocking, 5_CyclicL, 10473877, Sigl2022, 9136991, guin2018robust, yang2021looplock, Nuzzo2021, karmakar2019efficient} that increase the time complexity of attacks. Obfus-Lock \cite{ObfusLock} is proposed to leverage the skewness of nodes to construct a locked circuit and obfuscate the circuit using re-write rules. Furthermore, a theoretical method has been proposed to achieve both high query complexity and key error rates based on quasi-universal circuits, including convolutional biased target circuits \cite{zhou2019resolving}. In addition, recently, a sequential obfuscation solution called STATION \cite{STATION} has been proposed by leveraging disjoint encoding and combinational logic locking techniques. A comprehensive overhead and security analysis of state-of-the-art logic locking methods is also done in \cite{MLOverheadAnalysisofLogicLocking}. \textit{Despite the mentioned efforts, single-key solutions remain susceptible once the key is compromised, endangering the entire security of the hardware IPs.}

\subsection{Oracle-Guided Attacks}
\label{Sec:Attacks}
Boolean SAT solvers are used to reveal the correct key of logic-locked circuits using an oracle (i.e., an activated IC bought off the market) and a locked netlist to prune out the wrong key values \cite{ Rezaei:BreakUnroll, 8714924, 8342086, FunSAT}. The SAT-based attack \cite{evaluatingLogicEncryptionAlgorithms} uses Distinguishing Input Patterns (DIPs) that are specifically designed to exploit the discrepancies between the locked circuit and the oracle by targeting and identifying incorrect key values. The more incorrect key values the SAT solver eliminates in one iteration, the faster the attack can find the correct key. Then on, each attack has been strategically designed to target a specific defense mechanism; for example, Double DIP \cite{DoubleDIP} is used for attacking ICs locked with SAR-Lock \cite{SARLock}, where using two DIPs instead of one helps find the correct key faster. AppSAT \cite{AppSat} uses an approximate flow to find the probably-approximate-correct key in Anti-SAT \cite{AntiSAT} method. Fa-SAT \cite{Fa-SAT} inserts a single stuck-at fault at each signal of the locked circuit iteratively to find the correct key of BLE \cite{rezaei2020rescuing}. \textit{The assumption in all the above attacks is that there is a single static key in the logic-locked circuit to be deciphered.}

\subsection{Multi-Key Approaches} 
\label{Sec:MultiKey}
Recent works have brought the possibility of multi-key solutions. Specifically, DK-Lock \cite{DKLock} is a sequential locking method where one must provide two keys to a circuit; the first key is the activation key, which must be provided for a constant amount of time to activate the circuit, and then a functional key right after. DK-Lock may be susceptible to unrolling attacks \cite{Rezaei:BreakUnroll} that can expand the key size to reverse the method back to a single-key solution. SLED \cite{SLED} is another multi-key sequential solution but requires latches that operate on a clock, introducing additional complexity for combinational circuits. In addition, it depends on a seed value (i.e., a primary key) to operate, which can eventually be reduced to a single-key model since the attacker only needs to find out the seed value. \textit{Both of the mentioned multi-key logic locking methods may still be reverted back to a single-key model and thus susceptible to traditional SAT-based oracle-guided attacks. In addition, they depend on sequential components to be implemented.} 

Another multi-key logic locking solution, Gate-Lock \cite{GateLock}, uses an approach focused on locking gates, resulting in circuits that are resilient to SAT attacks. In Gate-Lock, the truth table has a height of $2^{n+k}$ while our proposed method maintains the same input size height of $2^n$ and only locks the outcomes within the truth table that are true. This allowed us to implement a more efficient algorithm.

\section{Multi-Key Logic Locking}
\label{Sec:Contribution}
In this section, after explaining the terminology, we discuss our proposed methods of locking a circuit with multiple keys; the first one locks the whole circuit, and it needs to generate a truth table for all the input combinations on the circuit, which may not be efficient in terms of space and time complexity. The second method, called \textbf{K-Gate Lock}, is a derivation but more optimized than the first one to focus on encoding the input combinations of the gates with specific key values. We also thoroughly discuss the implementation of the \textbf{K-Gate Lock} and evaluate the theoretical time complexity for any future oracle-guided attack to find the correct keys. An implementation example of \textbf{K-Gate Lock} on the c17 circuit from ISCAS 85 benchmarks \cite{iscas85} is shown in Figure \ref{fig:k-gate-locking-explanation}. You may refer to Table \ref{tab:resulst_algorithm_test} which contains the sequence of keys necessary to operate this locked circuit.
        
\subsection{Terminology} 
\label{Sec:Terminology}
In the context of \textbf{K-Gate Lock}, it is crucial to understand the terminology used to describe the various components and concepts:
\begin{description}
\item[n:] The total number of inputs to the original circuit.
\item[g:] The maximum number of gates to be locked within a circuit.
\item[k:] The number of inputs to a gate is often called the level of locking.
\item[gate key:] Each gate in a locked circuit has a specific key that controls its operation based on the input combination.
\item[key bit:] The individual binary elements that constitute a key.
\item[m:] The total number of bits in a key, aggregated from all key bits associated with each locked gate.
\item[keys:] Our approach uses keys derived from gate key combinations, with the specific key depending on the input.
\end{description}

\subsection{Locking the Whole Circuit}
The brute force method of locking a combinational circuit using multiple keys requires expanding the logic table of all the circuit's possible input/output combinations. Multiple keys can be inserted into each input/output combination when all the inputs and outputs are expanded. The truth table will maintain a size of $2^n$ since it does not create every combination of keys; it only adds the desired keys to an input combination. This brute force approach to locking a whole circuit is impractical because it would lead to an exponential-size truth table, and the implementation of locking a circuit would be time-consuming. For example, the C432 circuit in ISCAS 85 produces a truth table of 68,719 million rows, which is extremely large for one of the smallest benchmarks in ISCAS 85 suite. This motivated us to introduce \textbf{K-Gate Lock}.  

\begin{figure*}[]
    \includegraphics[width=0.97\textwidth]{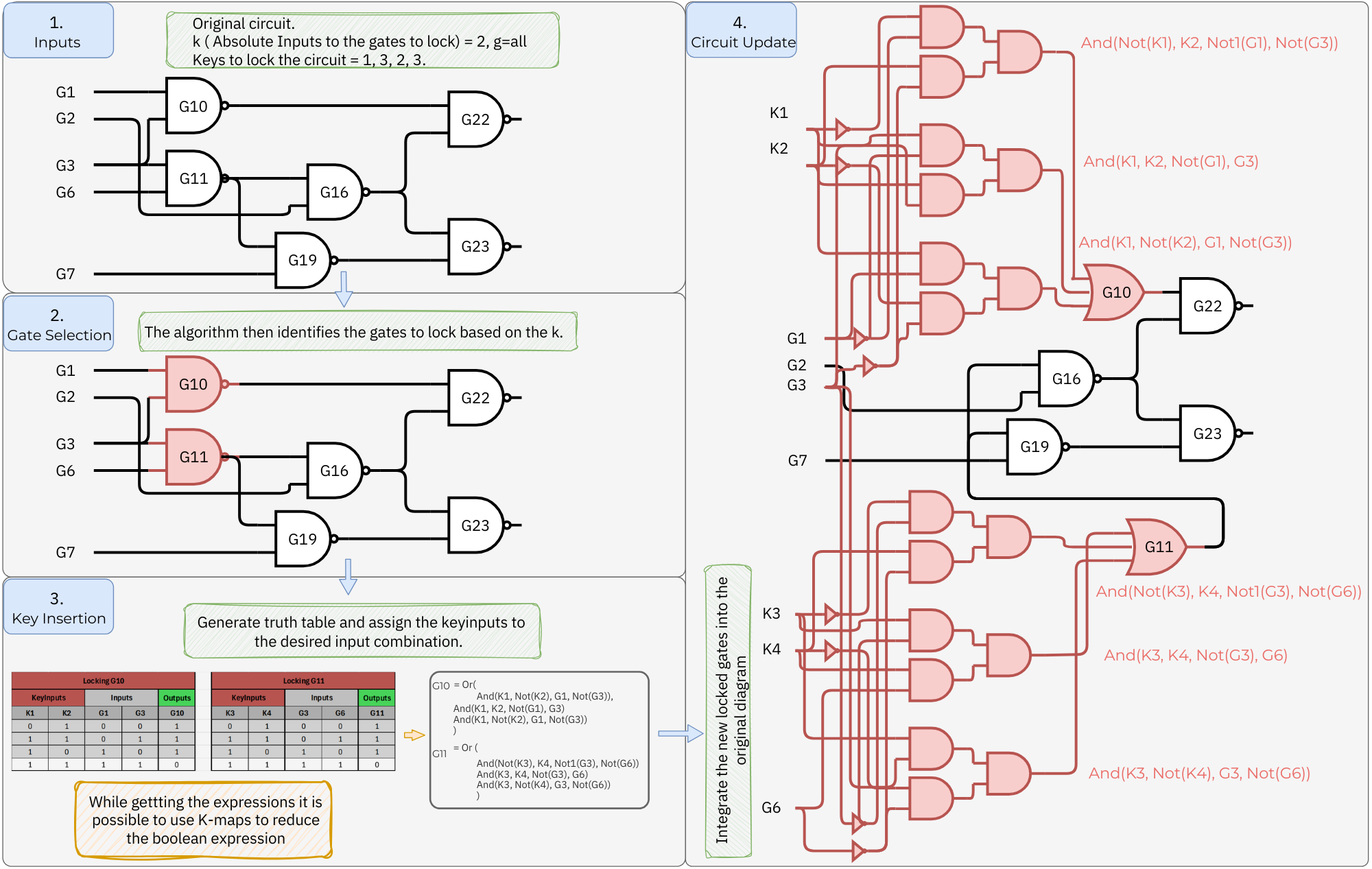}
    \caption{K-Gate Lock Example}
     \label{fig:k-gate-locking-explanation}
\end{figure*}
        
\subsection{K-Gate Lock Algorithm}
\textbf{K-Gate Lock} operates by locking specific gates within the circuit rather than the entire IC. This method utilizes the truth table of a gate or a more complex expression (i.e., a deep gate), encoding key bits directly into it. To operate the circuit, the user must provide a combination of inputs along with the corresponding keys in the correct sequence. 

Now, we discuss the steps for locking a circuit based on the circuit example in Figure \ref{fig:k-gate-locking-explanation}, which is locked at $k=2$ with the key values of $01$, $11$, $10$, and $11$.
       
\bluehighlight{\textbf{1. Inputs:}} The algorithm requires an input of the original circuit, unique keys, the maximum number of gates to lock (i.e., $g$), and a chosen level of gate locking (i.e., $k$). It is worth noting that the height of the truth table must be greater than the number of keys.
           
\bluehighlight{\textbf{2. Gate Selection:}} In the second phase of the process, the main task is determining which gates to lock, using the variable $k$ as a guide. The gates needed to be locked are those that have the number of absolute inputs equal to the input value $k$. For example, while locking the C17 benchmark shown in Figure \ref{fig:k-gate-locking-explanation}: 
\begin{itemize}
    \item G10 has two absolute inputs ($k=2$), G1 and G3.
    \item G11 has two absolute inputs ($k=2$), G3 and G6.
    \item G19 contains three absolute inputs ($k=3$), which are G7, G3, and G6.
    \item G16 contains three absolute inputs ($k=3$), which are G2, G3, and G6.
\end{itemize}

Since the initial constraint is to lock at $k=2$, only G10 and G11 are selected.

\bluehighlight{\textbf{3. Key Insertion:}} Up to this point, the algorithm has selected what gates to lock. In this step, the keys are inserted into the gate logic.  
For the C17 benchmark, as shown in Figure \ref{fig:k-gate-locking-explanation}, truth tables are expanded for the expressions $\overline{(G1 \land G3) }$ and $\overline{(G3 \land G6) }$, and the key bits are added as defined in the initial contains.

For deeper gates (i.e., $k>2$), it is necessary to have the absolute input to know when each key should be applied. It is also possible to maintain the original connections to the gate by using the inputs to the logic gate and adding them to the truth table. After the keys are added to the truth table, the expression can be extracted and simplified using any simplification method, like Karnaugh maps.

\begin{table}[!h]
    \centering
    \caption{Circuit Operation of Figure \ref{fig:k-gate-locking-explanation}}
    \setlength{\tabcolsep}{5pt} % Default is 6pt
    \begin{tiny}
    \begin{tabular}{|cccc|ccccc|cc|cc}
        \toprule
        k1 & k2 & k3 & k4 & G1 & G2 & G3 & G6 & G7 & G22 & G23  \\ 
        \midrule
        0 &  1 & 0 &  1 &  0 &  0 &  0 &  0 &  0  & 0 & 0  \\ 
        0 &  1 & 0 &  1 &  0 &  0 &  0 &  0 &  1  & 0 & 1 \\ 
        0 &  1 & 1 &  1 &  0 &  0 &  0 &  1 &  0  & 0 & 0 \\ 
        0 &  1 & 1 &  1 &  0 &  0 &  0 &  1 &  1  & 0 & 1 \\ 
        1 &  1 & 1 &  0 &  0 &  0 &  1 &  0 &  0  & 0 & 0 \\ 
        1 &  1 & 1 &  0 &  0 &  0 &  1 &  0 &  1  & 0 & 1 \\ 
        1 &  1 & 1 &  1 &  0 &  0 &  1 &  1 &  0  & 0 & 0 \\ 
        1 &  1 & 1 &  1 &  0 &  0 &  1 &  1 &  1  & 0 & 0 \\ 
        0 &  1 & 0 &  1 &  0 &  1 &  0 &  0 &  0  & 1 & 1 \\ 
        0 &  1 & 0 &  1 &  0 &  1 &  0 &  0 &  1  & 1 & 1 \\ 
        0 &  1 & 1 &  1 &  0 &  1 &  0 &  1 &  0  & 1 & 1 \\ 
        0 &  1 & 1 &  1 &  0 &  1 &  0 &  1 &  1  & 1 & 1 \\ 
        1 &  1 & 1 &  0 &  0 &  1 &  1 &  0 &  0  & 1 & 1 \\ 
        1 &  1 & 1 &  0 &  0 &  1 &  1 &  0 &  1  & 1 & 1 \\ 
        1 &  1 & 1 &  1 &  0 &  1 &  1 &  1 &  0  & 0 & 0 \\ 
        1 &  1 & 1 &  1 &  0 &  1 &  1 &  1 &  1  & 0 & 0 \\ 
        1 &  0 & 0 &  1 &  1 &  0 &  0 &  0 &  0  & 0 & 0 \\ 
        1 &  0 & 0 &  1 &  1 &  0 &  0 &  0 &  1  & 0 & 1 \\ 
        1 &  0 & 1 &  1 &  1 &  0 &  0 &  1 &  0  & 0 & 0 \\ 
        1 &  0 & 1 &  1 &  1 &  0 &  0 &  1 &  1  & 0 & 1 \\ 
        1 &  1 & 1 &  0 &  1 &  0 &  1 &  0 &  0  & 1 & 0 \\ 
        1 &  1 & 1 &  0 &  1 &  0 &  1 &  0 &  1  & 1 & 1 \\ 
        1 &  1 & 1 &  1 &  1 &  0 &  1 &  1 &  0  & 1 & 0 \\ 
        1 &  1 & 1 &  1 &  1 &  0 &  1 &  1 &  1  & 1 & 0 \\ 
        1 &  0 & 0 &  1 &  1 &  1 &  0 &  0 &  0  & 1 & 1 \\ 
        1 &  0 & 0 &  1 &  1 &  1 &  0 &  0 &  1  & 1 & 1 \\ 
        1 &  0 & 1 &  1 &  1 &  1 &  0 &  1 &  0  & 1 & 1 \\ 
        1 &  0 & 1 &  1 &  1 &  1 &  0 &  1 &  1  & 1 & 1 \\ 
        1 &  1 & 1 &  0 &  1 &  1 &  1 &  0 &  0  & 1 & 1 \\ 
        1 &  1 & 1 &  0 &  1 &  1 &  1 &  0 &  1  & 1 & 1 \\ 
        1 &  1 & 1 &  1 &  1 &  1 &  1 &  1 &  0  & 1 & 0 \\ 
        1 &  1 & 1 &  1 &  1 &  1 &  1 &  1 &  1  & 1 & 0 \\ 
        \bottomrule
    \end{tabular}
    \end{tiny}
    \label{tab:resulst_algorithm_test}
    \end{table}
    
\bluehighlight{\textbf{4. Circuit Update:}} In this step, the algorithm updates the original circuit with the new locked gates. 
As shown by step 4 of Fig~\ref{fig:k-gate-locking-explanation}, the locked G10 and G11 are placed instead of the original gates. Gates can be connected to the absolute inputs or the original inputs.

\bluehighlight{\textbf{5. Circuit Operation:}} To operate the locked circuit, one must input the correct keys used to encode the gates. For the example shown in Figure \ref{fig:k-gate-locking-explanation}, the truth table is shown in Table \ref{tab:resulst_algorithm_test}, where G1, G2, G3, G6, and G7 are the inputs and G22 and G23 are the outputs.

The \textbf{K-Gate lock} implementation is shown in Algorithms \ref{alg:lockCircuit} and \ref{alg:lockGate}; it relies on two functions: one to lock an individual gate and another to lock the entire circuit. In other words, the main focus of Algorithm \ref{alg:lockCircuit} is to find the gates to lock based on the given $k$ and replace the gates with the locked ones generated from Algorithm \ref{alg:lockGate}. The main focus of Algorithm \ref{alg:lockGate} is to lock an individual gate with a given set of dynamic keys. To attach the keys, it is necessary to generate the truth table of the gate's absolute inputs and then attach the given dynamic key.

\begin{algorithm}[!t]
\caption{Circuit Locking}
\label{alg:lockCircuit}
\begin{algorithmic}[1]
    \setlength{\labelsep}{15pt}  % Adjust this value to increase or decrease the space
    \STATE \textbf{Input:} Circuit $f(x)$, keys, g (max gates to lock), level $k$ (default $k = 2$)
    \STATE \textbf{Output:} Locked circuit $h(x, k)$
    \STATE $gates\_k\_inputs \leftarrow get\_gates\_with\_k\_inputs(f(x), k, g)$
    \STATE $h(x, k) \leftarrow f(x)$  %\COMMENT{copy the circuit to make modifications}
    \FOR{each gate in $gates\_k\_inputs$}
        \STATE $locked\_gate \leftarrow lock\_gate\_with\_key(gate, keys)$
        \STATE $h(x, k).replace(gate, locked\_gate)$
    \ENDFOR
    \RETURN $h(x, k)$ 
    \end{algorithmic}
\end{algorithm}

\begin{algorithm}[!t]
\caption{Gate Locking}
\label{alg:lockGate}
\begin{algorithmic}[1]
    \setlength{\labelsep}{15pt}  % Adjust this value to increase or decrease the space
    \STATE \textbf{Function} lock\_gate\_with\_key (gate, keys)
    \STATE \textbf{Input:} $gate$, $keys$
    \STATE \textbf{Output:} $locked\_gate$
    \STATE $table \leftarrow generate\_logic\_table(gate.inputs, gate.output)$
    \STATE $locked\_gate \leftarrow \text{()}$
    \FOR{each $(inputs, output)$ in $table$}
       \FOR{each $key$ in $keys$}
            \STATE $logic\_combination \leftarrow ((key + inputs) \rightarrow output)$
            \STATE $locked\_gate \leftarrow logic\_combination$
        \ENDFOR
    \ENDFOR
    \RETURN $locked\_gate$
    \STATE \textbf{End Function}
\end{algorithmic}
\end{algorithm}

\subsection{Time Complexity}
Now, we explain the time complexity of \textbf{K-Gate Lock}, which depends on the following:

\textbf{Number of Gates to Lock \((\min\{\frac{n}{k}, g\})\):} This number represents how many gates the algorithm aims to lock, determined by $g$ and $k$. The gates are chosen based on the level or absolute inputs they handle, as specified by $k$. This is demonstrated by line 3 of Algorithm \ref{alg:lockCircuit}, which identifies the gates to be locked. 

\textbf{Gate Locking Complexity \((2^k)\):} The locking mechanism for each gate involves encoding the gate key into the gate's logic. This requires creating a truth table for the gate with all possible combinations of inputs, resulting in \(2^k\) combinations. This step is shown in line 4 of Algorithm \ref{alg:lockGate}, which generates the truth table for the gate, and line 6, which encodes the gate key into the input combination.
   
Considering the above, the total time complexity can be represented as \(O(\min\{\frac{n}{k}, g\} \times 2^k)\). It is practical to fix $k$ at 2, leading to a linear time complexity for locking a circuit.   

\subsection{Attack Analysis} 
For attack analysis, we show how the time complexity increases for SAT-based oracle-guided attacks to find the correct keys in a circuit locked with \textbf{K-Gate Lock}. We explore the idea of traditional single-key SAT solvers and future multi-key SAT attacks that are aware of the \textbf{K-Gate Lock} method. 

\textbf{Single-Key SAT Attack:} Traditional SAT-based oracle-guided attacks are configured to find one correct key for a given circuit. Theoretically, such solvers are not suitable for finding multiple keys of the \textbf{K-Gate Lock} and in the best scenario, they will end up finding the first key of the sequence. We evaluate this with experimental results in Section \ref{Sec:Experiments}.
          
\textbf{Multi-Key SAT Attack:} In this case, the attacker is aware the circuit is locked with \textbf{K-Gate lock} and needs to explore the keys for every input combination as follows: \\ 
I) All input combinations: \textbf{$2^n$} possibilities (where $n$ is the number of inputs) as shown by the height of Table \ref{tab:timeComplexityTable}. \\
II) All potential values for each key: \textbf{$2^m$} possibilities (where m is the number of key bits). As shown in Table \ref{tab:timeComplexityTable}, the size of keys depends on the number of gates locked and the size of each gate key. The circuit designer has control over the total number of bits used and the number of gates. The $m$ parameter that depends on the number of locked gates $g$ and the key size for each gate is determined by the following equation:
             $$ m=\sum_{i=1}^{g} |key(i)|$$
                
Although current SAT-based attacks implement optimizations to prune several values of the global key, these optimizations cannot be applied to \textit{K-Gate Lock} because it uses a different key at every DIP. Consequently, SAT-based attacks are forced to perform a brute-force search for every key, resulting in a time complexity of $O(2^{m+n})$.

\begin{table}[!t]
\centering
\caption{Number of Keys for the Height of $2^n$ and $g$ Gates}
\setlength{\tabcolsep}{3pt} % Default is 6pt
\begin{footnotesize}
\begin{tabular}{c|ccccccc|ccc}
    \toprule
    & $k_1^1$ & ... & $ k_1^{|key_1|}$& . . . . &$k_g^1$ & ...  & $k_g^{|key_g|}$ & $inp_1$ & ... & $inp_{n}$   \\
    \midrule
     key 1      & x   &  ...    & x                  & . . . . & x      &  ...  &x                   &  0     &  ...  &  0              \\
     key 2      & x   &  ...    & x                  & . . . . & x      &  ...  &x                   &  0     &  ...  &  1              \\
     key 3      & x   &  ...    & x                  & . . . . & x      &  ...  &x                   &  0     &  ...  &  0              \\
     .          & .   &  ...    & .                  & . . . . & .      &  ...  &.                   &  .     &  ...  &  .              \\
     .          & .   &  ...    & .                  & . . . . & .      &  ...  &.                   &  .     &  ...  &  .              \\
    key $2^n$  & x   &  ...    & x                  & . . . . & x      &  ...  &x                   &  1     &  ...   &  1              \\
    \bottomrule
\end{tabular}
\end{footnotesize}
\label{tab:timeComplexityTable}
\end{table}

\section{Experimental Results} %untill here
\label{Sec:Experiments}
We conduct experiments on a Windows 11 machine, which accesses Linux Ubuntu 22.04 via WSL2. The machine is a Ryzen 7940HS with 8 cores and 16 threads at 4.0 GHz and 32 GB of DDR5 RAM. The source codes and created benchmarks of \textbf{K-Gate Lock} are publicly available on our GitHub repository\footnote{https://github.com/cars-lab-repo/KGL}. 

\subsection{Algorithm Validation}
The algorithm validation is done in Python along with pyEDA \cite{pyeda}, a tool used for electronic design automation. In this case, 2 gates of the c17 benchmark are locked with 4 different key combinations: $01$, $11$, $10$, and $11$. The truth tables are generated for the whole circuit shown in Table \ref{tab:resulst_algorithm_test} along with adding the correct key values; in this case, the original circuit and the locked circuit output the same value when the correct keys are fed in. 

Another set of tests is done using Netlist Encryption and Obfuscation Suite (NEOS) \cite{neos}, where circuits from ISCAS 85 \cite{iscas85} are locked using multiple keys that remain the same value: $101$, $101$, $101$. When multiple constant keys are provided, the SAT-based attack \cite{evaluatingLogicEncryptionAlgorithms} is able to find out the correct key, meaning it is also possible to achieve the level of locking based on the traditional logic locking methods. The results are shown in Table \ref{tab:static_key_benchmark_results}.
        
\begin{table}[!t]
\centering
\caption{Attack Results on Benchmarks with Static Keys}
% \begin{small}
\begin{tabular}{|c|c|c|c|c|c|}
    \hline
    \textbf{Benchmark}  & \textbf{ Gates} & \textbf{Time (S)} & \textbf{Reported Key}\\ \hline
    iscas85/c1355 &   3 & 0.08091 & \cellcolor{green!90} 101101101  \\ \hline
    iscas85/c17 &   2 & 0.01133 & \cellcolor{green!90} 101101  \\ \hline
    iscas85/c1908 &   3 & 0.1377 & \cellcolor{green!90} 101101101  \\ \hline
    iscas85/c3540 &   3 & 1.576 & \cellcolor{green!90} 101101101  \\ \hline
    iscas85/c432 &   3 & 0.03317 & \cellcolor{green!90} 101101101  \\ \hline
    iscas85/c499 &   3 & 0.04950 & \cellcolor{green!90} 101101101  \\ \hline
    iscas85/c5315 &   3 & 0.9743 & \cellcolor{green!90} 101101101  \\ \hline
    iscas85/c7552 &   3 & 1.942 & \cellcolor{green!90} 101101101  \\ \hline
    iscas85/c880 &   3 & 0.06435 & \cellcolor{green!90} 101101101  \\ \hline
\end{tabular}
% \end{small}
\label{tab:static_key_benchmark_results}
\end{table}

\begin{figure*}[]
\centering
\subfloat[Power (W)]
    {\includegraphics[width=\columnwidth]{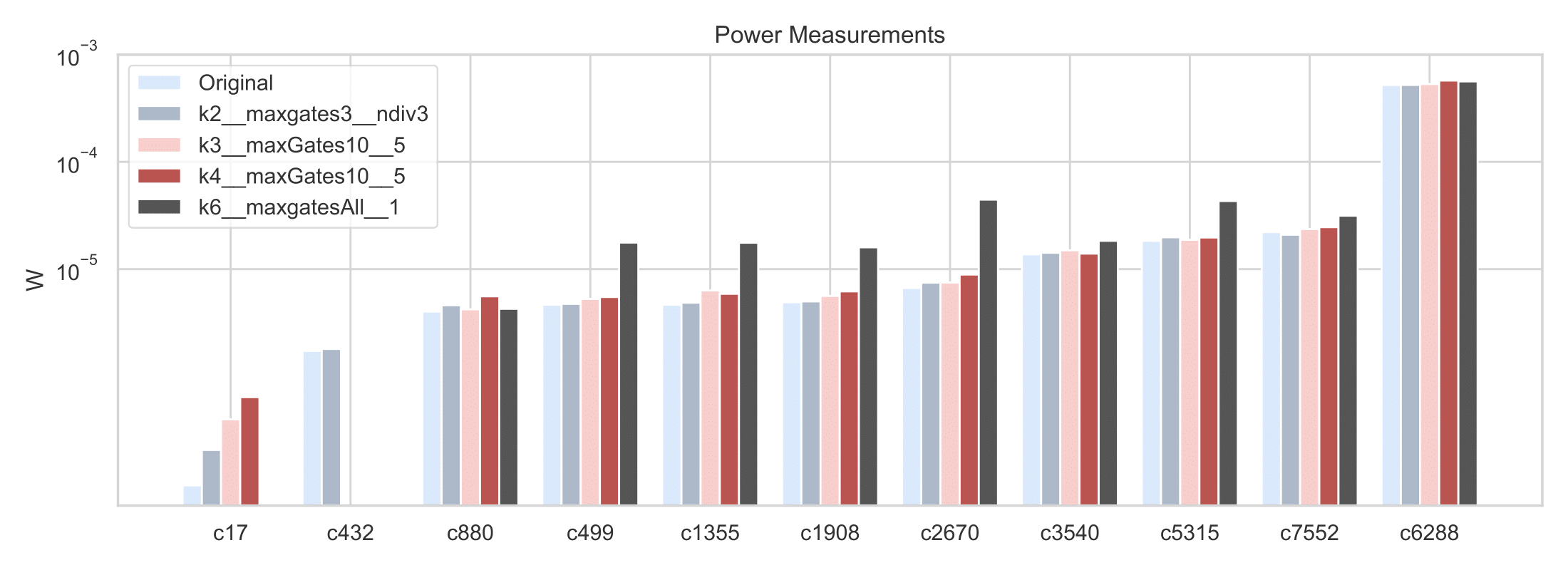}}
\subfloat[Number of Cells]
    {\includegraphics[width=\columnwidth]{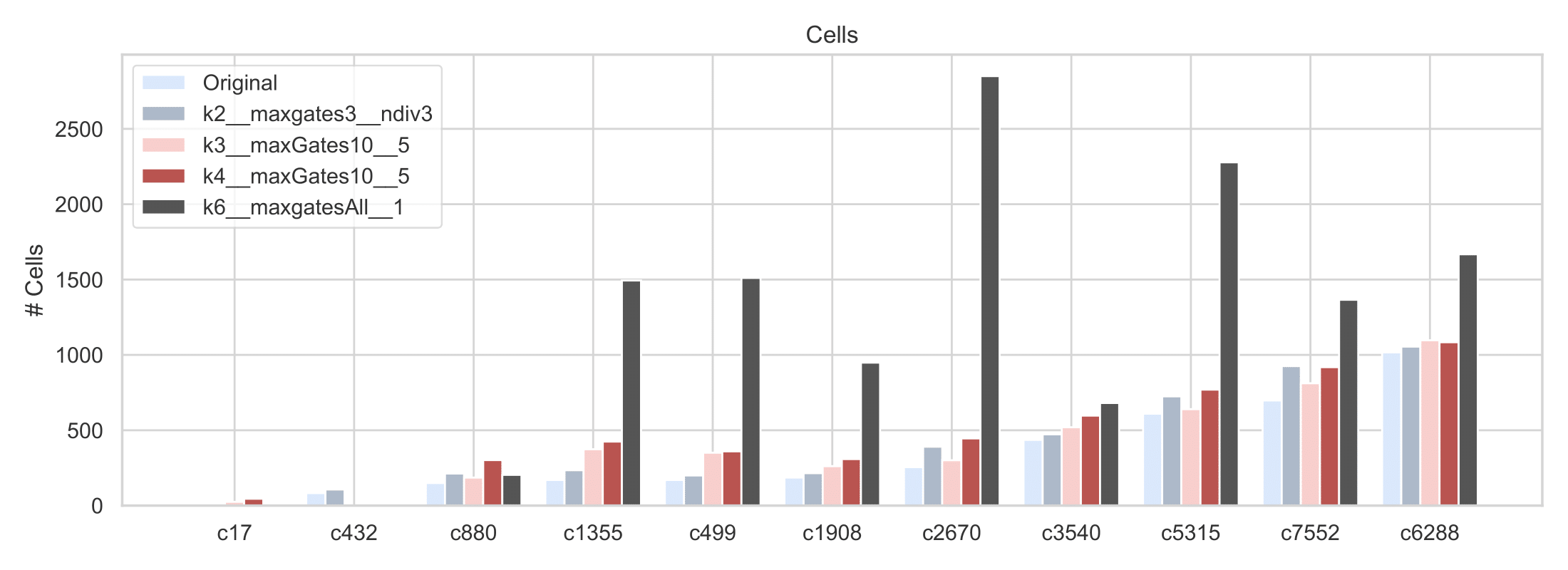}} \\
\subfloat[Area ($\mu m^2$)]
    {\includegraphics[width=\columnwidth]{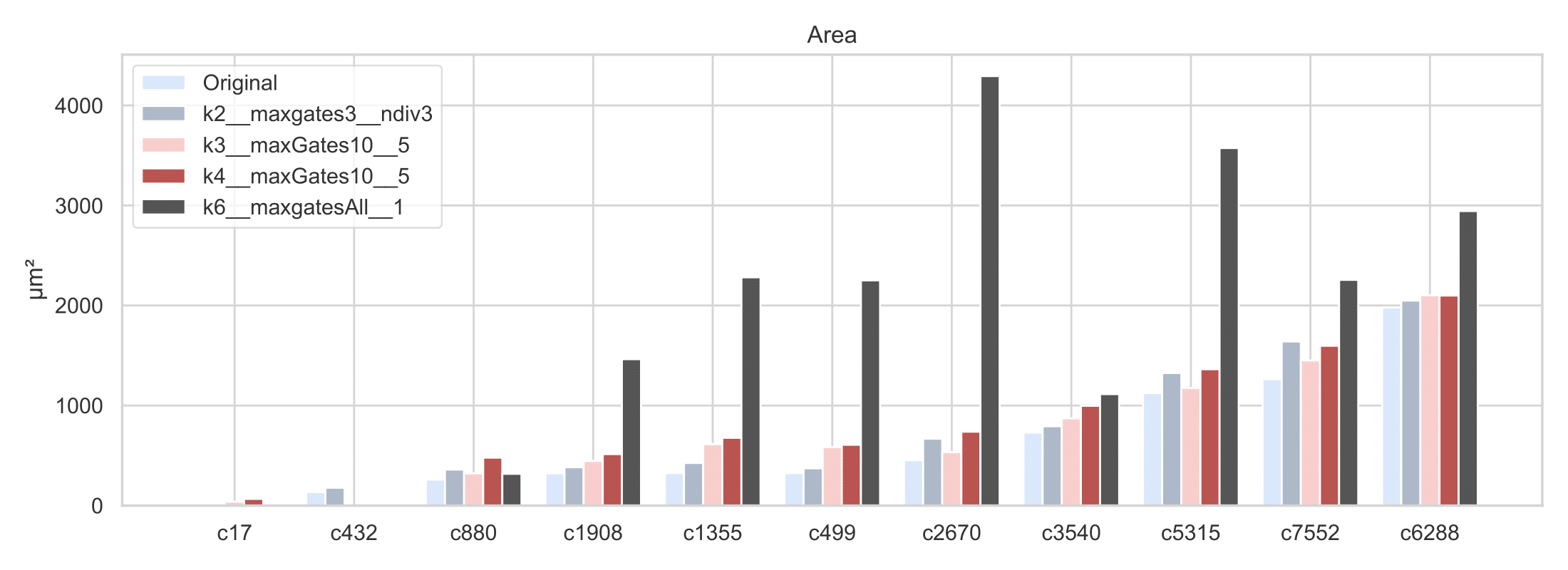}}
\subfloat[Number of IOs]
     {\includegraphics[width=\columnwidth]{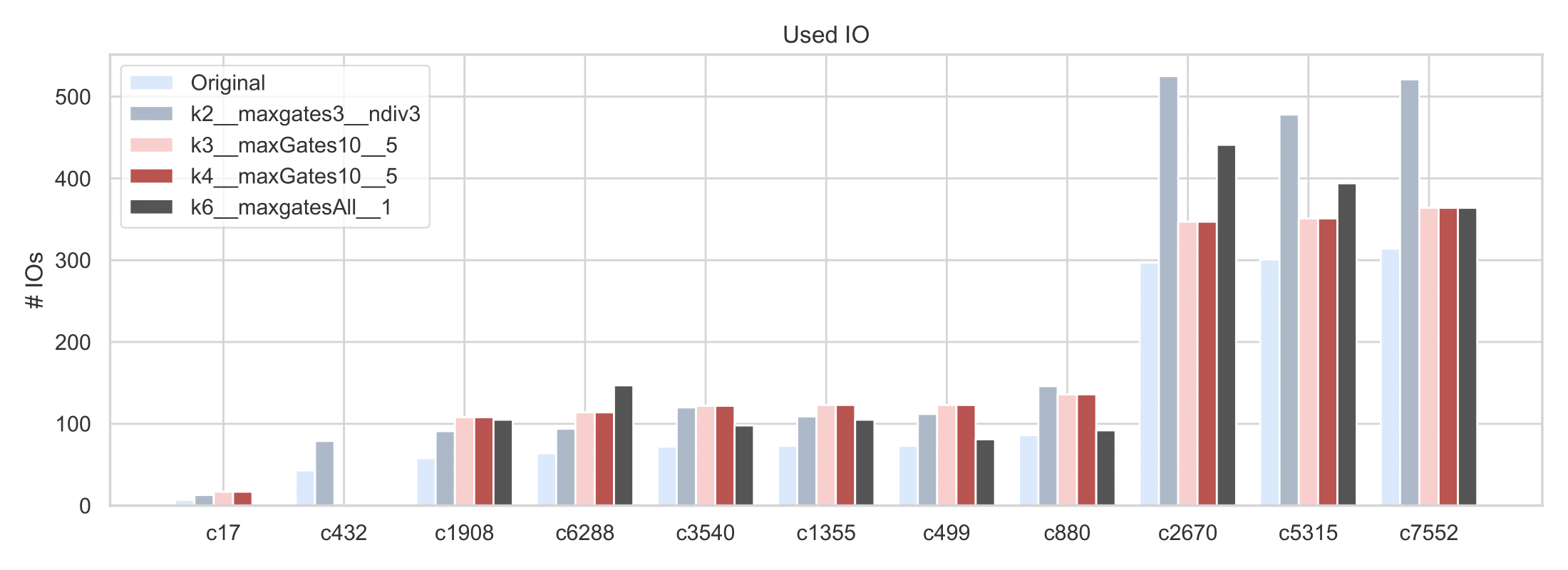}}
\caption{Overhead Measurements}
\label{fig:Original circuit vs encrypted circuit benchmarks}
\end{figure*}

\subsection{Security Evaluation}
The main objective of \textbf{K-Gate Lock} is to generate a locking mechanism that powerful SAT-based oracle-guided attacks will not be able to decrypt. We use combinational benchmarks of ISCAS 85 \cite{iscas85} and EPFL Benchmarks \cite{EPFL_benchmarks} as well as sequential benchmarks of ISCAS 89 \cite{iscas89}. Even though our proposed solution is based on combinational circuits, it is also possible to lock sequential circuits, locking portions of the circuit before including the flip-flops.
        
\subsubsection{Three Dynamic Keys} Now, we perform SAT-based oracle-guided attacks against the benchmarks locked with \textbf{K-Gate Lock}. We use NEOS \cite{neos} and RANE \cite{RANE} tools to run the attacks. The encryption for each circuit is done in \textit{.bench} files with our Python implementation of \textbf{K-Gate Lock}. 

The goal of this experiment is to demonstrate that with even a minimal number of keys, SAT-based attacks are unable to determine the correct keys. This limitation comes from their inherent design, which is to find only one key. The results  are shown in Table \ref{tab:test_results} in which the circuits are locked with the following gate key values: 
        \begin{itemize}
            \item  b'011 -  decimal value 3 
            \item  b'100 -  decimal value 4 
            \item  b'101 -  decimal value 5 
        \end{itemize}
        
\begin{table}[!t]
\centering
\begin{footnotesize}
\caption{ Attack Results on Benchmarks with 3 Small Dynamic Keys}
\begin{tabular}{|c|c|c|c|c|c|}
\hline
\textbf{Benchmark} & \textbf{Gates} & \multicolumn{2}{c|}{\textbf{NEOS}} & \multicolumn{2}{c|}{\textbf{RANE}} \\ \cline{3-6}
 &  & \textbf{Reported Key} & \textbf{Time (S)} & \textbf{Key Found} & \textbf{Time (S)} \\ \hline
iscas85/c1355 &  3 & \cellcolor{red!50}100100100 & 0.0941482 & \cellcolor{red!10} CNS & 1.27 \\ \hline
iscas85/c17 &  2 & \cellcolor{red!50}100101 & 0.0132606 & \cellcolor{red!50}011100 & 0.08 \\ \hline
iscas85/c1908 &  3 & \cellcolor{red!50}101011011 & 0.179116 & \cellcolor{red!50}101100101 & 0.69 \\ \hline
iscas85/c3540 &  3 & \cellcolor{red!50}011011011 & 1.95018 & \cellcolor{red!50}011011011 & 0.75 \\ \hline
iscas85/c432 &  3 & \cellcolor{red!50}011100100 & 0.0301723 & \cellcolor{red!50}011100100 & 0.14 \\ \hline
iscas85/c499 &  3 & \cellcolor{red!50}101101101 & 0.052921 & \cellcolor{red!50}100100101 & 0.36 \\ \hline
iscas85/c5315 &  3 & \cellcolor{red!50}011011101 & 0.462587 & \cellcolor{red!50}011011100 & 1.20 \\ \hline
iscas85/c6288 &  3 & \cellcolor{red!50}100100101 & 0.383951 & \cellcolor{red!50}101101100 & 2.86 \\ \hline
iscas85/c7552 &  3 & \cellcolor{red!50}011100100 & 1.95771 & \cellcolor{red!50}011011011 & 1.39 \\ \hline
iscas85/c880 &  3 & \cellcolor{red!50}101011100 & 0.0654231 & \cellcolor{red!50}101011100 & 0.25 \\ \hline
iscas89/s1196 &  3 & \cellcolor{red!50}011101011 & 0.081063 & \cellcolor{red!50}011101011 & 0.48 \\ \hline
iscas89/s15850 &  3 & \cellcolor{red!50}110011011 & 0.398465 & \cellcolor{red!50}000011101 & 55.53 \\ \hline
iscas89/s5378 &  3 & \cellcolor{red!10}CNS & 0.398465 & \cellcolor{red!50}010000011 & 11.03 \\ \hline
iscas89/s641 &  3 & \cellcolor{red!50}111000011 & 0.040502 & \cellcolor{red!50}000000010 & 4.71 \\ \hline
iscas89/s713 &  3 & \cellcolor{red!50}101100101 & 0.040502 & \cellcolor{red!50}000000101 & 4.22 \\ \hline
iscas89/s832 &  3 & \cellcolor{red!50}010010010 & 0.342658 & \cellcolor{red!50}010010011 & 1.53s \\ \hline
\end{tabular}
\label{tab:test_results}
\end{footnotesize}
\end{table}

In the tables, different colors are used to indicate specific conditions. The color light red\footnotemark[1] represents the ``Condition Not Solvable'' status. A deeper red\footnotemark[2] signifies a wrong key, while the darkest red\footnotemark[3] indicates that the attack failed.  Finally, green\footnotemark[4] denotes that the correct key has been found.

\begin{table}[!t]
\centering
\caption{Attack Results on Benchmarks with Dynamic Key Sizes Scalable to the Input Sizes}
\begin{small}
        \begin{tabular}{|c|c|c|c|c|}
        \hline
        \textbf{Benchmark} & \textbf{Key Size} & \textbf{NEOS  
        Time (S)} & \textbf{RANE  Time (S)} \\ \hline
        iscas85/c1355 & 40 & \cellcolor{red!50} 0.104842 & \cellcolor{red!50} 3.08 \\ \hline
        iscas85/c17 & 2 & \cellcolor{red!50} 0.0322511 & \cellcolor{red!50} 0.05 \\ \hline
        iscas85/c1908 & 30 & \cellcolor{red!50} 0.12339 & \cellcolor{red!50} 0.47 \\ \hline
        iscas85/c3540 & 50 & \cellcolor{red!50} 0.149944 & \cellcolor{red!50} 0.52 \\ \hline
        iscas85/c432 & 30 & \cellcolor{red!50} 0.164 & \cellcolor{red!50} 0.14 \\ \hline
        iscas85/c499 & 40 & \cellcolor{red!50} 0.151909 & \cellcolor{red!50} 0.45 \\ \hline
        iscas85/c5315 & 170 & \cellcolor{red!50} 2.14995 & \cellcolor{red!50} 3.46 \\ \hline
        iscas85/c6288 & 30 & \cellcolor{red!50} 0.894312 & \cellcolor{red!50} 2.77 \\ \hline
        iscas85/c7552 & 200 & \cellcolor{red!50} 1.90459 & \cellcolor{red!50} 1.32 \\ \hline
        iscas85/c880 & 60 & \cellcolor{red!50} 0.123312 & \cellcolor{red!50} 0.22 \\ \hline
        EPFL/adder & 60  & \cellcolor{red!50} 0.18 & \cellcolor{red!90} FAIL\\ \hline
        EPFL/bar & 50  & \cellcolor{red!50} 0.69 & \cellcolor{red!90} FAIL\\ \hline
        EPFL/div & 50  & \cellcolor{red!50} 461.62 & \cellcolor{red!90} FAIL\\ \hline
        EPFL/hyp & 60  & \cellcolor{red!50} 688.26 & \cellcolor{red!90} FAIL\\ \hline
        EPFL/log2 & 20  & \cellcolor{red!50} 25.73 & \cellcolor{red!90} FAIL\\ \hline
        EPFL/max & 80  & \cellcolor{red!50} 1.46 & \cellcolor{red!90} FAIL\\ \hline
        EPFL/multiplier & 50  & \cellcolor{red!50} 550.99 & \cellcolor{red!90} FAIL\\ \hline
        EPFL/sin & 20  & \cellcolor{red!50} 2.19 & \cellcolor{red!90} FAIL\\ \hline
        EPFL/sqrt & 50  & \cellcolor{red!50} 7.88 & \cellcolor{red!90} FAIL\\ \hline
        EPFL/adder\_depth\_2023 & 60 & \cellcolor{red!50} 0.9180 & \cellcolor{red!90} FAIL\\ \hline
        EPFL/arbiter\_depth\_2022 & 60 & \cellcolor{red!50} 0.9101 & \cellcolor{red!90} FAIL\\ \hline
        EPFL/bar\_depth\_2015 & 50 & \cellcolor{red!50} 12.12 & \cellcolor{red!90} FAIL\\ \hline
        EPFL/cavlc\_depth\_2022 & 10 & \cellcolor{red!50} 0.2943 & \cellcolor{red!90} FAIL\\ \hline
        EPFL/div\_depth\_2023 & 50 & \cellcolor{red!50} 152.8 & \cellcolor{red!90} FAIL\\ \hline
        EPFL/adder\_size\_2022 & 60 & \cellcolor{red!50} 1.224 & \cellcolor{red!90} FAIL\\ \hline
        EPFL/arbiter\_size\_2023 & 60 & \cellcolor{red!50}  0.2725 & \cellcolor{red!90} FAIL\\ \hline
        EPFL/bar\_size\_2015 & 50 & \cellcolor{red!50}  0.2725 & \cellcolor{red!90} FAIL\\ \hline
        EPFL/cavlc\_size\_2023 & 50 & \cellcolor{red!50}  0.63578 & \cellcolor{red!90} FAIL\\ \hline
        EPFL/div\_size\_2023 & 50 & \cellcolor{red!50}  83.98 & \cellcolor{red!90} FAIL\\ \hline
\end{tabular}
\end{small}
\label{tab:extended_test_results_detailed}
\end{table}

\begin{noindent}
    \begin{minipage}[!b]{\linewidth} %

  \vspace{0.5mm}   
\noindent\rule[0.5ex]{0.75\linewidth}{0.4pt} % Horizontal line, 75% of text width

    \footnotemark[1] \lightredhighlight{\phantom{x}} \cellcolor{red!10}CNS, \hspace{2mm}
    \footnotemark[2] \deeperredhighlight{\phantom{x}} \cellcolor{red!50}x..x,   \hspace{2mm}
    \footnotemark[3] \darkestredhighlight{\phantom{x}} \cellcolor{red!90}FAIL,   \hspace{2mm}
    \footnotemark[4] \greenhighlight{\phantom{x}} \cellcolor{green!100}Equal  \hspace{2mm}
\end{minipage}
\end{noindent}
        
This experiment shows that SAT-based attacks are not able to find the sequences of the keys but only to find the first one. The key found by the SAT solvers is a combination of the keys for the locked gates, as we see in the first test for Table \ref{tab:test_results}, the key value $100$ is repeated three times, which is only one of the combinations of the keys that are used to lock the gates.

\subsubsection{Dynamic Key based on Input Size}
For the second security experiment shown in Table \ref{tab:extended_test_results_detailed}, we explore scaling key sizes to input sizes. Specifically, for ISCAS 85 benchmarks, the gate key size is calculated using $\left\lfloor \frac{n}{10} \right\rfloor$ with 10 gates locked, resulting in a floor value of 10 compared to the input size. In addition, we use a logarithmic scaling formula to deal with the high number of inputs in the EPFL benchmarks. While the EPFL benchmarks \cite{EPFL_benchmarks} are in \textit{.blif} format, we use the ABC tool \cite{abc} to convert them to \textit{.bench} files and perform the attack. The key values are generated randomly within the range dictated by the key input size for both benchmark suites. To simplify the testing process, we limit the number of gates locked to 10. This aims to approximate the size of the key as closely as possible to the input size. The experimental results highlight the challenges faced by SAT-based attacks in thwarting dynamic key locking. 

\subsection{Overhead Analysis}
Now, we analyze the overhead of \textbf{K-Gate Lock}. The experimental setup utilizes Cadence Genus, using low mapping and optimization effort. Circuit locking is executed using \textit{.bench} files and these are converted to Verilog with the ABC tool \cite{abc}. Power measurements, as shown in Figure \ref{fig:Original circuit vs encrypted circuit benchmarks}, are performed on various ISCAS 85 \cite{iscas85} benchmarks with the following locking parameters. 
        \begin{itemize}
          \item \textbf{Test Run 1 }: $k=2$, $g=3$, $key \ bit=\frac{n}{3}$
          \item \textbf{Test Run 2 }: $k=3$, $g=10$, $key \ bit=5$
          \item \textbf{Test Run 3 }: $k=4$, $g=10$, $key \ bit=5$
          \item \textbf{Test Run 4 }: $k=6$, $g=all$, $key \ bit=1$
        \end{itemize}

As shown in Figure \ref{fig:Original circuit vs encrypted circuit benchmarks}, power consumption does not increase much, and the area increases only slightly, correlated with the number of inputs and the gates locked. 
The missing data in the area, I/O measurements, along with power and temperature, means that the circuit is unable to lock any gates given the specific specifications, leading to the failure of the locked circuit creation. %For example, the algorithm failed to lock the $C432$ circuit within the specified $k$-th level range in Test Run 3 because it does not have any gates that have 6 absolute inputs; it also failed to lock gates at levels 4 and 3 for the same reason. 
We calculated the average percentage increase using the sum of the whole test run and compared it with the original sum. These values are shown in Table~\ref{tab:summaryOverhead}. 
The highest jump in terms of power is 25.62\% in \textit{Test Run 4}, while the smallest jump is less than 1\% in \textit{Test Run 1}. 

\begin{table}[!t]
        \centering
         %\begin{footnotesize}
        \small 
        \caption{Average Overhead}
        \label{tab:summaryOverhead}
        \begin{tabular}{lrrrr}
            \toprule
                Test Name               &  Power \% &  Cells \% &  Area \%  &  I/O \% \\
            \midrule
             K=2, g=3, key bit=n/3       & 0.45     & 20.40         & 18.63        & 64.84    \\
             K=3, g=10, key bit=5 bits   & 2.93     & 23.07         & 19.52       & 33.14     \\
             K=4, g=10, key bit=5 bits   & 10.45     & 41.33        & 33.95        & 33.14     \\
             K=6, g=All, key bit=1 bit   & 25.62     & 246.33       & 197.89        & 35.23    \\
            \bottomrule
        \end{tabular}
         %\end{footnotesize}
    \end{table}
            \section{Conclusion}
\label{Sec:Conclusion}
In this paper, we proposed a novel multi-key logic locking solution called \textbf{K-Gate Lock} that is based on input encoding and can be fully implemented using combinational logic without the need for state-holder components. Experimental results showed that \textbf{K-Gate Lock} is resilient against state-of-the-art SAT-based oracle-guided attacks with minimal overhead. %In addition, we proved that any future advanced attack from this family that targets unlocking \textbf{K-Gate Lock} will forcefully end up with exponential time complexity. 
This offers the potential of multi-key logic locking schemes for robust hardware IP protection with reasonable overhead.

\section*{Acknowledgment}
This material is based upon work supported by the National Science Foundation under Award No. 2245247.

\end{document}